\documentclass[manuscript,table]{acmart}
\AtBeginDocument{%
  \providecommand\BibTeX{{%
    \normalfont B\kern-0.5em{\scshape i\kern-0.25em b}\kern-0.8em\TeX}}}

\setcopyright{acmcopyright}
\copyrightyear{2018}
\acmYear{2018}
\acmDOI{XXXXXXX.XXXXXXX}

\acmConference[CHI '25]{ACM Conference on Human Factors in Computing Systems}{April 26-- May 1,
  2025}{Yokohama, Japan}
\usepackage{multirow}
\usepackage{subcaption}
\usepackage{tabularx}

\usepackage[commandnameprefix=ifneeded,final]{changes}

\begin{document}

\title[The Shape of Agency]{The Shape of Agency: Designing for Personal Agency in \replaced{Qualitative Data}{Computational Thematic} Analysis}

\author{Luka Ugaya Mazza}
\affiliation{%
  \institution{University of Waterloo}
  \city{Waterloo}
  \country{Canada}}
\email{r3woo@uwaterloo.ca}

\author{Plinio Morita}
\affiliation{%
  \institution{University of Waterloo}
  \city{Waterloo}
  \country{Canada}}
\email{}

\author{James R. Wallace}
\email{james.wallace@uwaterloo.ca}
\orcid{0000-0002-5162-0256}
\affiliation{%
  \institution{University of Waterloo}
  \city{Waterloo}
  \state{Ontario}
  \country{Canada}
}

\renewcommand{\shortauthors}{Ugaya-Mazza et al.}

\begin{abstract}
Computational thematic analysis is rapidly emerging as a method of using large text corpora to understand the lived experience of people across the continuum of health care: patients, practitioners, and everyone in between. However, many qualitative researchers do not have the necessary programming skills to write machine learning code on their own, but also seek to maintain ownership, intimacy, and control over their analysis. In this work we explore the use of data visualizations to foster researcher agency and make computational thematic analysis more accessible to domain experts.  We used a design science research approach to develop a datavis prototype over four phases: (1) problem comprehension, (2) specifying needs and requirements, (3) prototype development, and (4) feedback on the prototype. We show that qualitative researchers have a wide range of cognitive needs when conducting data analysis and place high importance upon choices and freedom, wanting to feel autonomy over their own research and not be replaced or hindered by AI. 
\end{abstract}

\maketitle

\section{Introduction}


Qualitative research enables researchers to understand the meanings, experiences, and views of participants, giving voice to the lived experiences of people. \added{It provides a deep understanding of complex phenomena by exploring subjective experiences in rich detail \cite{denzin2011sage} and placing findings within their social, cultural, or situational contexts \cite{creswell2016qualitative}.} \deleted{For instance,} \citet{Rice1996} paints qualitative research as important for describing and explaining behaviour, understanding how people perceive their own health experience, and how to best design \replaced{for}{a health program that takes into consideration} those that will use it. \added{It can be particularly important when seeking to identify new variables or concepts that might otherwise be overlooked, while capturing emotions and subjective perspectives—dimensions often neglected in purely numerical data \cite{patton2014qualitative}.}

However, qualitative analysis \replaced{can also be}{is} labour-intensive, rigorous, and exhausting \cite{Jiang2021,Pope2000}. It can take weeks for researchers to manually curate and analyze data sets and \deleted{often} this process is \added{often} \deleted{further} complicated by a lack of available time \cite{Jiang2021,Feuston2021}. These challenges are further exacerbated when considering the vast amounts of data available to researchers wishing to engage with online sources like social networking sites (e.g. \cite{Gauthier2022,Gauthier2023}). \added{When considered together, these challenges mean that qualitative researchers are often faced with a lack of agency when working with the large data sets available today from sources like social media.}

Computational thematic analysis seeks to combine established qualitative research methods (e.g., \cite{Braun2006,boyatzis_transforming_1998}) with emerging techniques from machine learning (e.g., \cite{blei2003latent,NIPS2017_3f5ee243,devlin2018bert}). Human-computer interaction research has in turn sought to develop visual interfaces that make these techniques accessible to domain experts who may not have the programming expertise required to perform computational thematic analysis without technical support (e.g., \cite{Gauthier2022,Hong2022}). However, recent findings suggest that domain experts are unlikely to use tools that they do not trust; they are hesitant to give up intimacy, ownership, and agency in their own analyses \cite{Jiang2021,Feuston2021}.

%
%
In this work we explore domain experts' experiences around using data visualizations as a tool for knowledge translation for computational thematic analysis. Data visualizations can be useful when conducting larger scale research, since they allow for pattern recognition, data exploration, summariz\replaced{ation}{ing}, and \replaced{external cognition}{cognitive assistance} \cite{Braga2020, Gomes2019, Cabitza2016, Kennedy2020,Jiang2021, Hong2022}.\deleted{Especially when it comes to AI, there is a benefit to using data visualizations during research to assist with finding patterns and connections.} We engaged in design science research to understand their perceptions of \emph{personal agency} --- which is the positive feeling related to being given the space to act and see those actions shape the world \cite{Murray1997, Eichner2014}  --- when working with large data sets, and with computational tools and data visualizations to support their analyses. Our approach included an iterative design exercise which was informed by a close reading of the literature and qualitative data analysis software, interviews with domain experts, low-fidelity prototyping, and follow-up interviews with the same experts to provide feedback on those prototypes. 

%
%

Our results show that personal agency can be used as a \added{design} lens for \deleted{developing} software interfaces\deleted{, especially when trying to bridge the gap between human-computer interaction}. Personal agency \deleted{as a design lens} provides a novel way of understanding how specific design features can foster agency and thus assist in incorporating them into the interfaces that attend to \replaced{users'}{the} needs \deleted{of the users}. It is particularly useful when considering human-AI interactions, \replaced{where loss of agency has been shown to be a primary concern of end users \cite{Jiang2021}.}{since works like Jiang et al. [34] point to the loss of agency as a primary concern for qualitative researchers to adopt the use of AI in their research.}


%
%
In summary, we make the following contributions: 

\begin{enumerate}
    \item Show how \emph{personal agency} as conceptualized by \citet{Eichner2014} can be used as a design lens in HCI
    
    \item Investigate qualitative researchers' perceptions of how AI can support their work, particularly within the context of thematic analysis of large data sets.
    
    \item We engaged in a co-design exercise in which qualitative researchers helped to design a series of novel, low-fidelity prototypes
    
    \item We reflect on barriers and facilitators to the use of AI in qualitative research, and how qualitative researchers' own perceptions of AI shifted during the co-design exercise.  
\end{enumerate}



\section{Related Work}

The human-computer interaction research community has recently explored the use of social media as a resource to understand people. The insights generated can be quite valuable, for instance, the ability to post on social media anonymously has been found to support disclosure around sensitive topics, such as addiction and recovery \cite{IWNDWYT}, disordered eating \cite{Nova2022}, parenting issues \cite{momdadit}, intimate partner violence \cite{rodriguez2020computational}, and mental health \cite{nazanin,Xu2023}. These disclosures can be absent in data collected using methods like interviews or surveys since people may hesitate to speak about sensitive or stigmatized issues \deleted{in interviews} for privacy reasons or fear of being judged \cite{momdadit}.

However, a significant barrier to performing qualitative analysis on social media data is its scale, which requires researchers to sample data for topics of interest. While in the past, researchers have focused on randomized samples, modern computational techniques enable for \emph{purposive} sampling over large data sets through unsupervised topic modelling \cite{IWNDWYT,Gauthier2022}. In exploring the use of these techniques, the human-computer interaction research community has noted `convergences' \cite{muller_machine_2016,baumer2017comparing} between qualitative and machine learning research: they are both grounded in data, interpretive, and iterative in nature. And in further pursuing these convergences, they have explored both technological advances and visual interfaces to support their use by non-programmer domain experts. 

Recent work in HCI has in particular explored the use of AI and ML techniques to support qualitative analysis.  For instance \citet{dai2023llmintheloop} developed LLM-enhanced qualitative analysis techniques. Similarly, \citet{lennon2021developing} created AQUA, an automated qualitative analysis assistant. CollabCoder \cite{gao2023collabcoder} and \citet{Xiao2023} further explore the use of LLMs to support \emph{codebook} thematic analysis, with an emphasis on use of codes and inter-coder reliability. Further, \citet{Chen2018} examines how LLMs can be used to highlight ambiguity, and co-construction of models. However, while these advances provide qualitative researchers with an unprecedented ability to explore, search, and understand data from social media there is also a growing body of research pointing to barriers to adoption. For instance, \citet{Jiang2021} and \citet{Feuston2021} have recently shown that qualitative researchers have reservations about the use of automated tools in the context of qualitative data coding. 

Interest has therefore turned to creating novel visual interfaces for these techniques. For instance, \citet{klein2015exploratory} proposes visual techniques for exploratory data analysis. Similarly, ClioQuery \cite{Handler2022} provides a comprehensive visual interface for exploring historical documents, and leverages natural language processing techniques to support human understanding of trends over time. The Computational Thematic Analysis toolkit \cite{Gauthier2022,Gauthier2023} and Scholastic \cite{Hong2022} also provide visual interfaces for non-programmers to engage with advanced ML techniques.

\added{Despite these advances in visual interfaces, research has pointed to lack of agency as a barrier to adoption.} People seldom prefer full automation \replaced{over}{, tending towards} delegation \deleted{with humans leading a task} \cite{Feuston2021, Jiang2021, Lubars2022}. Having humans at the centre of decision making ensure\added{s} that researchers feel autonomy, ownership, and agency over their own projects \cite{Jiang2021}. \added{These recent findings in HCI motivated our exploration of personal agency as a way of ensuring that qualitative researchers maintain their feelings of autonomy in their research.}

\added{The HCI community has made significant strides in studying agency, but there remains substantial controversy surrounding how agency should be measured and understood, as noted by \citet{Bergström2022}. Research in this area generally falls into two categories: explicit agency, where users have direct, visible control over actions and outcomes \cite{Tajima2022, Kasahara2021}, and implicit agency, where users experience agency subtly, often through system behaviours that guide or support their actions \cite{Coyle2012a,Didion2024}. \citet{Bennett2023} provides a comprehensive overview of this work over the past 30 years, and finds that HCI has overwhelmingly focused on aspects of agency like self-causality, but that ``it was rare for authors to explicitly articulate particular aspects of agency and autonomy, let alone discuss coordination, tensions, and trade-ofs between them.'' In this work we therefore examine designing for agency through \citet{Eichner2014} theoretical lens.}


\deleted{These alternative systems to incorporate AI into qualitative research are examples of delegating tasks between people and machines. Frameworks for the delegation of tasks to machines have existed long before the age of AI, for example, Parasuraman et al .[46] define four types of human-machine interaction and their possible levels of automation. The authors highlight automation can be beneficial when the task at hand is high in mental workload, does not remove situational awareness of the worker, does not incite complacency, or skill degradation, and can be reliable and low cost.}

\deleted{With the introduction of AI, best practices dictate that the users themselves choose when to use and what roles it should automate [40]. Lubars and Tan [40] defend that AI delegation should be based on motivation, perceived difficulty, perceived risk, and trust. Of these, trust is the element that most closely relates to agency.}

\deleted{Trust strongly governs whether people choose to delegate tasks or not, and people tend to delegate more to systems that they trust [38, 40]. Lubars and Tan [40] discuss trust in machine automation in terms of performance, process, purpose, and value alignment. This means that people tend to trust machines when they believe it will reach the goals set by them with transparency as to the process, as well as align to their personal values and needs [40].}

\deleted{Notably, to date, none of the interfaces mentioned have specifically designed for \emph{agency}. This raises questions about how we can foster agency through design and encourage the use of computational techniques in qualitative research, and how qualitative researchers envision using computational techniques on large-scale data. In this work we develop a better understanding of how domain experts  hope to use automation in their research process, through co-designing an interface for facilitating human-AI interaction.}

\subsection{Agency}

In Bandura's \cite{bandura2006toward} theory of agency, humans are not simply the products of their lives, but \replaced{active contributors to their experiences}{contributors to it}. By being agents of their own lives, \added{human} agency implies intention, forethought, self-\replaced{regulation}{reactivess}, and self-reflectiveness \cite{bandura2006toward}. This means \replaced{individuals}{people} have the ability \added{to} strategize, project a future from the present, adopt intentional plans accordingly, and then reflect upon their \replaced{actions and}{own} effectiveness. \added{The} same concept can be extrapolated to agency in media interaction, \replaced{where it fosters a sense of engagement}{which derives the same feeling though meaningful actions} and the consequences that \replaced{emerge from}{derive from} it \cite{Murray1997}. However, agency is not \replaced{experienced equally by}{equal for} everyone, as different environments \added{and contexts} will alter the perception of agency \cite{Eichner2014}.

Furthermore, \added{the execution of} an action itself is not necessary for agency to be perceived \cite{Murray1997}. People can \replaced{perceive}{still sense} agency even when their actions are limited \added{or constrained} (i.e., a game of chess). This \replaced{perception arises}{happens} because those limited actions have immediate consequences and cause change \cite{Murray1997}. The more a choice \replaced{affects}{changes} an environment, the more people experience agency, regardless of whether the choice \replaced{yields}{had} a positive or negative impact \cite{Murray1997}. When interacting with media, \added{agency involves not only action but} also the possibility of taking an action if so desired, and sometimes, this \replaced{sense of}{perceived} agency is just as powerful as action-taking \cite{Eichner2014}.

Complementing this concept, \citet{Tanenbaum2009} \replaced{defines}{define} agency as a commitment to meaning, emphasizing that it happens when a person \replaced{actively chooses}{chooses} to immerse \replaced{themselves}{into a narrative} and believes in what they are doing \cite{Tanenbaum2009}. Therefore, agency is not simply acting within a world, but \replaced{the expression of one’s intentions to act and the reception of}{expressing your intentions of acting and receiving} feedback from \replaced{the surrounding environment}{your surroundings} \cite{Eichner2014,Murray1997,Tanenbaum2009}.

\citet{Eichner2014} divides agency into three levels: personal, creative, and collective. Personal agency is fostered through the mastery of narrative, choice, action, and space and must be perceived before someone experiences creative and collective agency. Creative agency \replaced{arises}{comes} from the creation of material for the media one is interacting with—for example, how people create MODS for video games \cite{Tanenbaum2009} --- while collective agency is fostered through the creation of a community around a piece of media (e.g., Comic Con). In this work, we focus only on the four dimensions of personal agency since they must be perceived before the others:

\begin{itemize}

\item \textbf{Mastering of action} occurs when the user moves their body and sees the same movement reflected \replaced{in}{on} the media they are interacting with \cite{Murray1997}. For example, in the game A Ceremony of Innocence, the movement of the mouse on the screen acts as an extension of the human body, and the user perceives agency through its movement \cite{Bizzocchi2003}. In non-game contexts, a responsive mouse cursor \replaced{can}{elicits} elicit the same feelings.

\item \textbf{Mastering choice} \replaced{occurs when}{is when} the user realizes they can make a choice within their environment \cite{Eichner2014}. This type of agency works best when there are \replaced{structured rules constraining}{rules that restrict} the world around the user. \replaced{Conversely, when too much freedom exists, agency tends to be fostered instead by a commitment to meaning}{instead, when there’s too much freedom, agency is more likely to be fostered through a commitment to meaning} \cite{Tanenbaum2009}.

\item \textbf{Mastering narrative} happens when one identifies the genre of media they are interacting with and can predict what will happen next, \replaced{deriving satisfaction}{thus gaining pleasure} from seeing a narrative unfold \replaced{as anticipated}{in the way they predicted} \cite{Eichner2014}. For example, when viewing a line graph, the \replaced{graph’s progression}{lines} may act as a \replaced{narrative guide}{narration point} that guides the reader through the data \cite{luka2023}, which \replaced{enables a line graph to foster}{makes a line graph foster} greater agency than the same data in a table.

\item Lastly, \textbf{mastering of space} is when the user perceives agency through the act of moving through space \cite{Murray1997}. In order for this to happen seamlessly, the user needs to have enough knowledge of the media they are interacting with to use movement effectively. For example, if \replaced{a user attempts}{you try} to move and the zoom does not function in the way \replaced{they expect}{you are used to}, then the agency is \replaced{diminished, creating frustration instead}{traded for frustration} \cite{luka2023, Bizzocchi2003}.

\end{itemize}

Specific \replaced{to the development of}{for developing} data visualizations, agency is also fostered through the process of \emph{creation}, \replaced{where}{as in,} the user perceives their agency because they \replaced{actively co-create}{create} the visualization \replaced{in collaboration}{with} \replaced{with the designer via}{the designer through} interactions \cite{Rawlins2014}. It might be a simple zoom \replaced{action or}{,} an input of numbers, but since users can take actions that \replaced{impact}{have effects on} the visualization, agency is perceived through that mutual crafting of meaning \cite{Rawlins2014,Tanenbaum2009}. Additionally, the agency of the user is mediated by the agency of the designer, thus making the \replaced{final visualization product}{end product (visualization)} a collaboration among the designer, \replaced{the visualization, and}{visualization, and} the end user, each operating with their respective constraints and \replaced{contexts}{context} \cite{Rawlins2014}. When the user is given no control over the \replaced{visualization’s}{message, data, and} design, then they experience very limited agency. When they are given unconstrained control over these three elements, then their sense of agency \replaced{can overshadow}{overpowers} the designers’ \cite{Rawlins2014}.



\begin{table}[tb!]
\centering
\caption{Overview of Agency Types as Defined by \citet{Eichner2014}, \added{detailing Individual, Creative, and Collective Dimensions and their key characteristics in media interaction.}}
\begin{tabular}{l p{10cm}}
\toprule
\textbf{Type of Agency} & \textbf{Description} \\
\midrule
Personal Agency & \replaced{Focused on}{Rooted in} individual actions, perceptions, and decisions. \\
  \quad Mastering of Narrative &  \replaced{Recognizing}{Understanding} the genre and predicting outcomes based on \replaced{narrative}{contextual} cues. \\
  \quad Mastering of Choice & Exercising \replaced{one’s}{decision-making} capabilities \added{within defined constraints}. \\
  \quad Mastering of Action & \replaced{Perceiving one’s physical or digital}{Seeing one's real-world} actions reflected in media. \\
  \quad Mastering of Space & Navigating \replaced{through}{within} physical or digital \replaced{environments}{spaces} to achieve goals. \\
\\
Creative Agency & \replaced{Involves}{Emerges from} creating objects, content, or modifications that relate to and expand on the primary media. \\
\\  
Collective Agency &  \replaced{Built upon}{Derived from} interactions within a community that engages \replaced{collaboratively}{with the same media} and co-creates shared meanings.  \\
\bottomrule
\end{tabular}
\end{table}

\section{Overview of Research Process}

\begin{figure}[tb!]
    \centering
    \includegraphics[width=\textwidth]{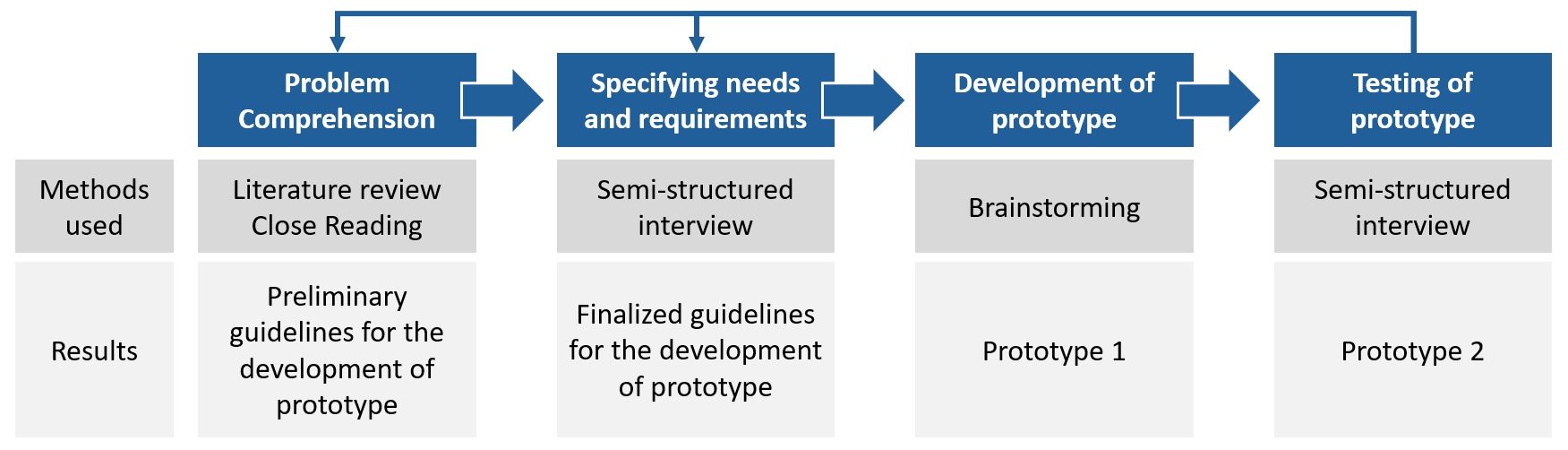}
    \caption{Our Design Science Research process followed a series of four phases: 1) Problem Comprehension, 2) Specifying user needs and requirements, 3) Development of prototype, and 4) Testing of prototype.}
    \label{fig:enter-label}
\end{figure}

Design Science Research (DSR) \cite{adam2021design,Maguire2001,Santos2018} is an iterative process that seeks to generate scientific knowledge through the creation of a design artifact for the solution of a specific problem. It consists of 5 overall phases: 1) planning the design science process, 2) \replaced{problem comprehension}{understanding the specific context of use}, 3) specifying \added{needs and} requirements, 3) \replaced{development of prototype}{producing prototypes}, and 5) testing the prototypes against requirements. \added{the cyclical nature of the method allows for iterativity between the phases, allowing the researchers to further understand the problem and modify requirements as they develop the solution.}

Design Science Research (DSR) has many parallels with Research Through Design, which has been more broadly adopted in the HCI community \cite{Zimmerman2007}. Both methods call for the development of an artifact to solve a specific problem and the creation of knowledge through that development, as well as an iterative design process with multiple instances of user feedback \cite{Santos2018,Zimmerman2007}. 


For our project, phase 1 was dedicated to further understanding the problem proposed through a literature review and analysis of other similar software out in the market through a technique called Close Reading \cite{fadel2020metodo,bizzocchi2011well, bizzocchi2005run}. Then, phase 2 was dedicated to understanding the specific needs and requirement of the users, which was achieved through a semi-structured interview with users, in this case qualitative researchers. \replaced{The results from both phases were then cross analyzed in order to develop the guidelines used for the development of the low fidelity prototype in phase 3}{With those phases completed, guidelines were developed for phase 3, the development of a low fidelity prototype}. Following that, phase \replaced{4}{3} was a beta-testing of the prototype in which the participants of the interview in phase 2 returned for another session of user input. This phase had the goal of both testing the prototype and confirming that the design choices that came out of the first interview were what the participants had intended. A visual representation of the entire method can be seen in \replaced{\autoref{fig:enter-label}}{Figure 4}. Problem comprehension and prototype development were performed independently by the research team, whereas the other phases were performed collaboratively with domain experts recruited to our study.

\section{Specifying Needs and Requirements}

We conducted semi-structured interviews to examine researchers' needs and wants for a tool to help with their qualitative data analysis. The goal of it was to determine \replaced{three}{3} aspects of data visualization interaction: 

\begin{enumerate}
    \item How do qualitative researchers use visual elements to assist in their data analysis? 
    \item What visual elements do they find important to have in a data visualization? 
    \item How do qualitative researchers perceive their sense of agency while doing research? 
\end{enumerate}

The interview was conducted with the purpose of understanding the user and task requirement \cite{Maguire2001}. The sample of participants was gathered via purposive/snowball sampling seeking information power \cite{Malterud2016}. \deleted{This project has a narrow aim since the situation being observed is highly specific. It also utilized previously established theories to conduct data collection and analysis, as well as be descriptive in nature.} The data was collected and analyzed by researchers that have experience in dealing with qualitative data, therefore the sample sought higher levels of information power instead of opting for a comprehensive large sample. \deleted{To guarantee more variety, the participants were purposefully selected to gather a wide array of experiences across different areas of health care.}



\subsection{Recruitment Criteria}
We recruited a total of five (5) qualitative researchers through snowball sampling at our local institution. All were in the field of healthcare, but had varied specialties, and different levels of qualitative research experience and experience with data visualizations. All participants had limited to no experience with programming, but different levels of interest and experience in using AI for their data analysis. 

\begin{table}[tb!]
    \centering
    \caption{Summary of participants' experience with qualitative research and data visualization.}
    \begin{tabular}{l l l l}
    \toprule
    Participant & Field of Study & Qualitative Exp. & Datavis Exp. \\
    \midrule
    1 & Vision Science & New & Moderate \\
    2 & Social Media & High & Moderate \\
    3 & Technology & High & High \\
    4 & Gerontology & Moderate & Low \\
    5 & Aging & Moderate & Low \\
    \bottomrule
    \end{tabular}
\end{table}




\subsection{Interview protocol}
The interview protocol consisted of eight open-ended questions which prompted participants to describe their process for conducting data analysis, the types of data visualization they are familiar with, and what software (if any) they use. \deleted{We asked participants to consider examples of different data visualization techniques from von Engelhardt [59];  maps, diagrams, tables, and charts, with each having subdivisions, for a total of 10 kinds of data visualization. Finally,} \added{We} collected sketches of participants' ideal data visualization tool, as well as examples of how they use data visualizations as cognitive assistance during their own data analysis process. 

\subsection{Analysis Method}
We used reflective thematic analysis to understand the interview data. Thematic Analysis is a qualitative method that consists of finding and refining themes within data to draw connections and groups (e.g., \cite{Braun2006,clarke2017thematic}). We conducted the thematic analysis deductively with the theories of fostering of agency. The interviews were transcribed using Microsoft Teams and Microsoft Word and then imported into NVIVO for the analysis. The main researcher read the transcripts several times to extract preliminary themes and then ran them over a secondary coder to further refine them. After the refining, a second analysis was run by the main researcher and the themes were finalized.

\subsection{Results}

\renewcommand{\thesubfigure}{\arabic{subfigure}}
\begin{figure}[tb!]
 \centering
 \begin{subfigure}{.49\textwidth}
        \centering
        \includegraphics[width=\textwidth]{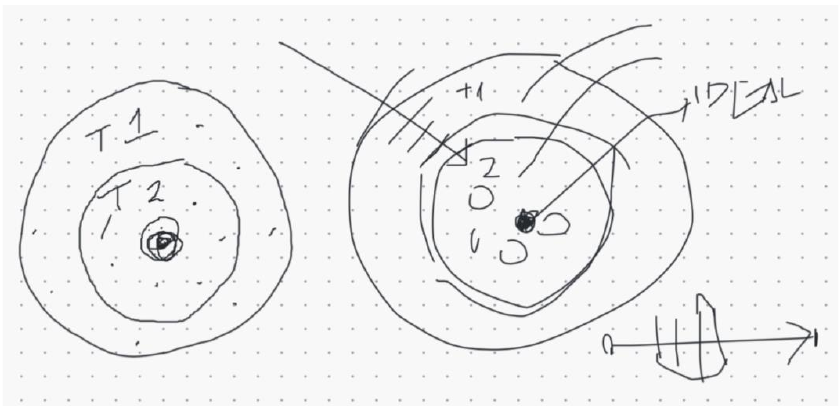}
        \caption{A grouping plot by Participant 1}\label{s1}
\end{subfigure}%
\hfill
\begin{subfigure}{.49\textwidth}
        \centering
        \includegraphics[width=\textwidth]{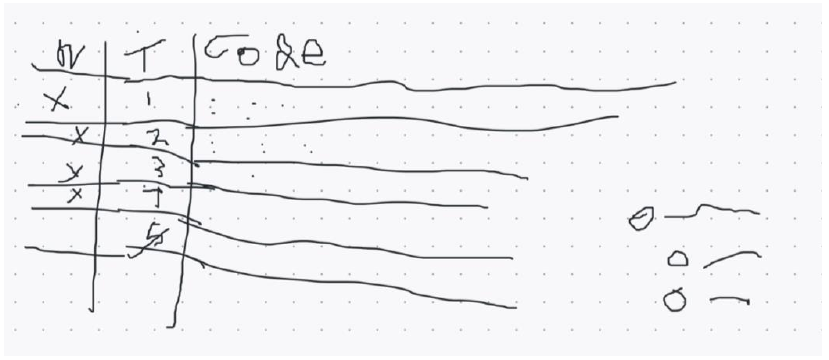}
        \caption{A grouping plot by Participant 2}\label{s2}
\end{subfigure}%
\vspace{.5cm}
\begin{subfigure}{.49\textwidth}
        \centering
        \includegraphics[width=\textwidth]{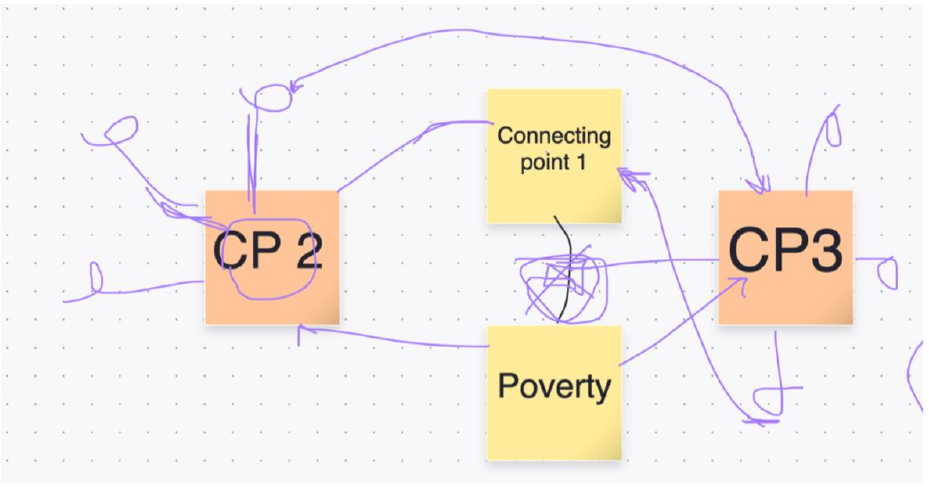}
        \caption{A mind map from Participant 4}\label{s3}
\end{subfigure}%
\hfill
\begin{subfigure}{.49\textwidth}
        \centering
        \includegraphics[width=\textwidth]{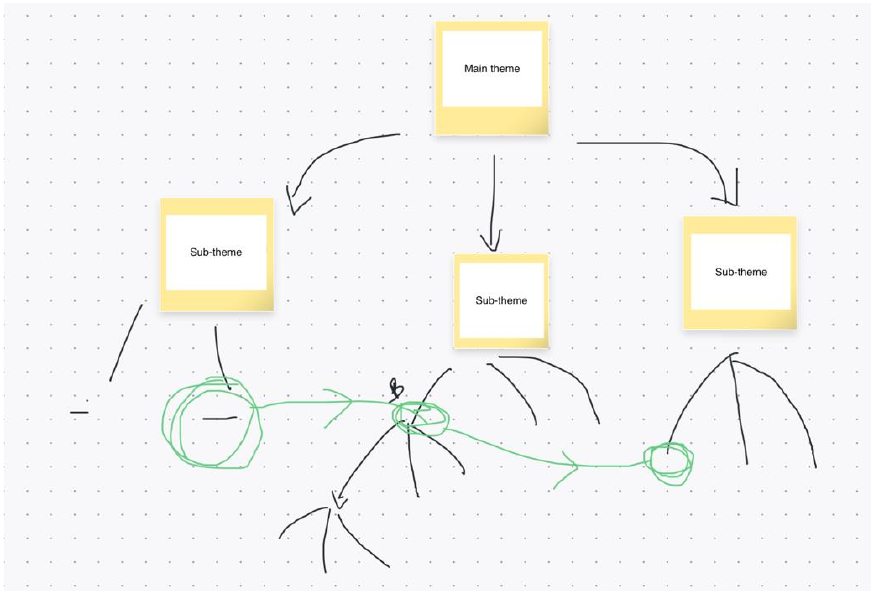}
        \caption{A mind map from Participant 5}\label{s4}
\end{subfigure}%
 \caption{Participant Sketches captured during our interviews. Participants used a variety of techniques to group (1,2) and mind map (3,4) during their qualitative research practices.}
 \label{fig:visualizations}
\end{figure}

Data visualizations were unanimously used as part of \replaced{participants'}{their} qualitative research process. Through their use, participants reported being able to think, process, and connect ideas. For example, \replaced{P}{p}articipant 4 shared (\autoref{fig:visualizations}) and explained how she likes to use a combination of a grouping diagram and a mind map to draw connections between themes and create clusters \deleted{that eventually lead to her being able to conduct her thematic analysis. She said ``That's how I get things out of my head so that I'm not just juggling them in my head so that they can be clear.''} \added{We therefore developed codes across three categories: agency, cognitive assistance, and visual literacy. Within each category we also developed sub-codes. Agency was divided into the four kinds of mastery, choice, narrative, action, and space. Visual literacy was divided into statements related to their awareness of how to represent data, their preferences for visualization, their experience in the datavis, and their ability for abstraction. Lastly, cognitive assistance was divided into assistance for oneself, or assistance for others.}

In discussing how they used visualization, participants described two general \replaced{uses}{categories}: supporting their own sense-making and cognition \deleted{(e.g., grouping, frequency, comparison, making connections, tracking changes over time, path, and seeing processes),} and sharing the results of that work with others \deleted{(e.g., presentation and summarization)}. Often visualizations supported more than one need at a time, for example, participant 5 said “I organize my thoughts in \deleted{like} a little notepad. I tried, for example, I try to make a story, like there should be, like a theme that makes sense, right?” which implies she organizes, explores, and searches for a narrative in her data all using the same visualization. As knowledge translation tools, participants highlighted the importance of visualization for presentation and summary. Participant 3 framed this need as “the story in action”, in which the researcher needs to present industry partners with visualizations that tell a narrative strong enough to enact change, answering questions like “what's the point? What can I do with it? How do we act going forward?”. 

\added{Participants expressed a desire for software that could better assist them during these two phases of qualitative research. We therefore focused our analysis on agency as a design lens during these activities.} \replaced{Overall,}{Finally,} in reflecting on the role of computational tools in supporting thematic analysis, participants expressed the importance of \replaced{three main factors on their perception of agency: trust, delegation, and guidance.} {three factors: trust, delegation, and facilitators of personal agency.} 

\deleted{Despite the reported use of data visualizations, some participants complained about being limited based on their lack of knowledge on what datavis exists. Participant 5 explained ``I think I'm restricted with like linking, you know, diagrams visualization cause that's how I have always envisioned qualitative research to be. Now that you're showing me so many types of visualization, I'm like oh damn! Does that even exist?'' after seeing a different way of representing data that she considered to be better suited for her data than the one she was previously using. This sentiment was mirrored by participant 1, who was very interested in knowing what would be the best way of showcasing her data, but unaware of how to discover new possibilities. She shared her excitement for a tool that could help with showing her what kind of data visualizations exist and how she can best use them.}

\deleted{Additionally, software like NVIVO was described as being too complicated to use for generating data visualizations. Participant 1 shared she felt intimidated by the platform's seemingly complicated interface and feeling frustration when trying to figure out how to make the kinds of data visualizations she wanted. Similarly, participant 2 shared that she dislikes NVIVO's interface and does not even bother learning it as she views the visualizations as not useful enough for the purposes she needs. On the other hand, excessive freedom was also quoted as being a deterrent for exploration, since the participants could not use data visualization to their full potential if they did not know what existed to be used. Lack of guidance reportedly caused frustration when participants were left to figure out how to visualize data with no idea on how to use the proper tools. Tools that did not provide sufficient guidance were also considered to be more frustrating than helpful.}

\subsubsection{Trust}


All participants voiced a lack of trust in a system that automates data analysis, fearing replacement or loss of autonomy. For instance, two participants feared being replaced by machines. Participant 4 \replaced{said}{expressed fears of becoming obsolete, quoting,} ``I wouldn't want it to analyze anything for me because then I become obsolete, so I don't want it to be too good at its job, where I become obsolete and it replaces me and takes over the world and it turns into the matrix.'' Conversely, four participants highlighted the importance of preserving context in qualitative data analysis, questioning a machine's ability to understand and generate codes without this understanding. Participant 1 called the use of automatic coding "very, very bizarre," doubting a computer's ability to grasp data nuances.

Similarly, Participant 4 explained that \deleted{the human element of qualitative analysis is very important and that} immersion in data is crucial to any qualitative analysis and cannot be replaced by machines. Participant 3 echoed the same feeling, saying that ``Naturally, still, you know, being a researcher and knowing the context, being able to read between the lines, there would still be a lot of things that you would need to code manually because you know in a way the machine doesn't know.'' She expressed her frustration with computational methods, calling them rudimentary and questioning whether they would even be useful for the kind of research she does, since her qualitative analysis is so context heavy and she does not believe an algorithm could pick up on these nuances. Participant 1 illustrated the same point by saying ``I didn't look at [NVIVO’s automatic coding system] in that great detail because the codes that it produced, I think they were all single word codes, which wasn't really helpful anyway''. 


\subsubsection{Delegation}

While not willing to completely trust machines to perform a qualitative analysis, all participants were open to delegating smaller tasks. Three participants were open to the idea of machine-supported analysis, but were very strict as to which tasks should be delegated and which should remain in human hands. Participant 1 shared how she enjoys having the machine assist in tasks such as counting, checking frequency, and helping her design data visualizations\replaced{, but which would only provide suggestions for more complex tasks which she could then decide on.}{. She expressed the need to have a tool that could assist with the last item especially, since her experience with data visualizations is limited.} \added{Participant 3 was openly receptive to the idea of qualitative coding because it enabled her to engage with larger data sets from social media. She said that machine assistance would not negate the need for a human researcher, but would enjoy delegating some of the work.}


\deleted{Similarly, Participant 3 shared the same ideal of ``cutting down on the work of coding a little bit.'' She explained that her ideal partnership would be using machine support as a facilitator for checking the frequency of very simple sentiments. Since she questions the abilities of machine learning to fully understand context, she would prefer to keep automatic coding to simple sentiments that machines would not have trouble processing, explaining that ``There's very simple sentiments that people express, like ‘technology is expensive’, ‘[name of devices] are expensive’. That's something that would be nice that you wouldn't have to manually tag like just those… those very simple, straightforward ideas.''}

\deleted{Participant 3 was more openly receptive to the idea of qualitative coding with machine assistance. She explained that because of the nature of her type of research in working with social media, her samples end up being very large and very short in character amount. Because of this, she explained that coding by hand is simply not possible and if she could employ the assistance of AI to make the process more seamless, it would be ideal. She shared ``I prefer [coding with the assistance of AI]. I don't know how to…. when it's 500,000 data points, it's not possible for me to do it. I can do it with 4000 easy peasy, but 500,000 plus. I just don't know if it's possible and so, having the toolkit which would. It can do that for you.'' She clarified that machine assistance would not negate the need for a human researcher, saying that she still codes herself, but enjoys delegating some of the work.} 

On the other hand, two participants shared their reluctance towards delegating at all. Participant 5 explained that she views qualitative research as untainted and would prefer it not ``be ‘invaded by machines.’ I still want it to be in the hands of humans.'' She explains that having a human be the one to derive codes to be the beauty of qualitative research and that she fears losing the empathy and personal connection gained through qualitative data analysis. Similarly, Participant 4 talked about her perceived sense of agency and how much it would be affected if she felt the machine was doing her work for her. She quotes ``I feel like a part of it is the agency element like I feel like. It'll replace me or like it does things very fast that like. You know, it's concerning.'' This statement highlights her sense of loss of agency, as well as her fear of being replaced. 


\subsubsection{\added{Guidance}}

\added{Despite all participants reporting use of data visualizations, some participants complained about being limited based on their lack of knowledge and awareness of different visualization idioms and techniques. Participant 5 explained ``I think I'm restricted with like linking, you know, diagrams visualization 'cause that's how I have always envisioned qualitative research to be. Now that you're showing me so many types of visualization, I'm like oh damn! Does that even exist?''. This sentiment was mirrored by Participant 1, who was interested in learning about how to showcase her data, but unaware of how to discover new possibilities. She shared her excitement for a tool that would support the creation of data visualizations.}

\added{Additionally, software like NVIVO was described as being too complicated to use. Participant 1 shared she felt intimidated by the platform's complicated interface and frustrated when trying to figure out how to make the data visualizations she wanted. Similarly, Participant 2 shared that she dislikes NVIVO's interface and views the visualizations as not useful enough for her purposes. On the other hand, excessive freedom was quoted as being a deterrent for exploration, since participants could not use data visualization to their full potential if they did not know what existed. Echoing Participant 1's comments around delegation, the need for suggestions and guidance was a consistent theme throughout the interviews, particularly in the context of how best to visualize data, like determining which kind of visualization to use based on current needs. Participant 4 described her ideal interface as ``laying those options out for me and giving me suggestions. I think that would be very helpful.'' Her ideal tool would guide her through the process of designing a data visualization, whilst still leaving her with enough choice to feel in control of the process.}



\deleted{After delegating certain tasks, participants still wanted to maintain ultimate control over decisions made during the analysis. Three participants discussed how they liked being presented with options when working with software, instead of being given un-editable results. Participant 1 illustrated this point by saying her ideal interaction would be through the machine offering insights and suggestions, which she could then sort through and choose to accept. She explained her preference as, ``I can see the different [suggestions] and then select the one that resonates with me.'' Similarly, participant 2 talked about liking the idea of an assistant that could suggest possible themes and connections, while still leaving her in charge of whether to keep them or not.} 

\deleted{The use of suggestions and guidance was a consistent theme through the interviews, particularly in the context of how best to visualize data, like determining which kind of visualization to use based on current needs. Participant 4 described her ideal interface as ``laying those options out for me and giving me suggestions. I think that would be very helpful.'' She wanted her ideal tool to present her with layout options and guide her through the process of designing a data visualization, whilst still leaving her with enough choice to feel in control of the process.}

\deleted{One potential source of this desire was a lack of experience and knowledge in the field of visualization, but also a recognition of its importance. Participant 1 shared her lack of experience with data visualizations and perceived lack of tech-savviness, and the belief that because of this she considered herself limited in choosing data visualizations and knowing how to best represent her data. Participant 5 echoed the same feeling, saying that because of her lack of expertise in data visualization she considers her skills to be restricted to only the diagrams that she knows. She explained ``I don't really know if there's like a specific… because I haven't really like, have experience with data visualization in particular.''}

\section{Development of Digital Paper Prototype}

Based on the results of our problem comprehension and specifying needs \added{and} requirements phases, we \deleted{next set out to} create\added{d} a low-fidelity prototype \deleted{that would enable us to further examine computational support for thematic analysis}. To help to focus that development, we decided on some guiding principles grounded \deleted{in what we had learned from the interviews} \added{in the concepts of personal agency}: 

\begin{enumerate}
    
    \item \textbf{Support data visualization based on researcher needs}. \deleted{Participants expressed interest in support for visualization based on their research goals.} \replaced{The software should}{To attend to that, we set out to create a visual interface that} \replaced{organize}{separated the different} data visualizations into \added{meaningful} tasks like: exploratory data analysis, frequency checking, making connections, comparing, tracking changes over time, grouping, seeing processes, and mapping out paths.  \added{This organization will support mastery of action by providing researchers with clear options that align with their needs, and provide them with the ultimate decisions about which to use.}
    

    \item \replaced{Moreover, the} {The} \replaced{software}{interface} should \textbf{provide \added{awareness and} guidance for what kinds of data visualization exist and how they could be used}. \replaced{P}{since p}articipants described \deleted{expressed an interest in using different kinds of data visualizations, but} a lack of knowledge of how to best use \replaced{visualizations}{them} to achieve their goals. \added{The use of guidance would serve to foster agency through choice, since it allows for researchers to explore their options freely, and then making decisions based on their own needs and understanding.}

    \item Give researchers \textbf{editing capabilities} so they can fine tune \deleted{the} visualizations to fit \replaced{their own}{whatever} narrative \deleted{they desire} (e.g., editing colours, moving elements, etc.) \added{to foster agency through building a visualization, designing it, and moulding the narrative behind it.}

    \item \textbf{Preserve autonomy} through showing the researcher how \deleted{the} results were reached. \deleted{To do that, our tool presents choices to the researcher, explains how results were reached, and allows the researcher to have  have the final say over everything.} \added{All work done by the computer should be transparent and give the user the final say to foster agency through choice and ensure their autonomy.}


\end{enumerate}

\begin{figure}[tb!]
    \centering
    \includegraphics[width=\textwidth]{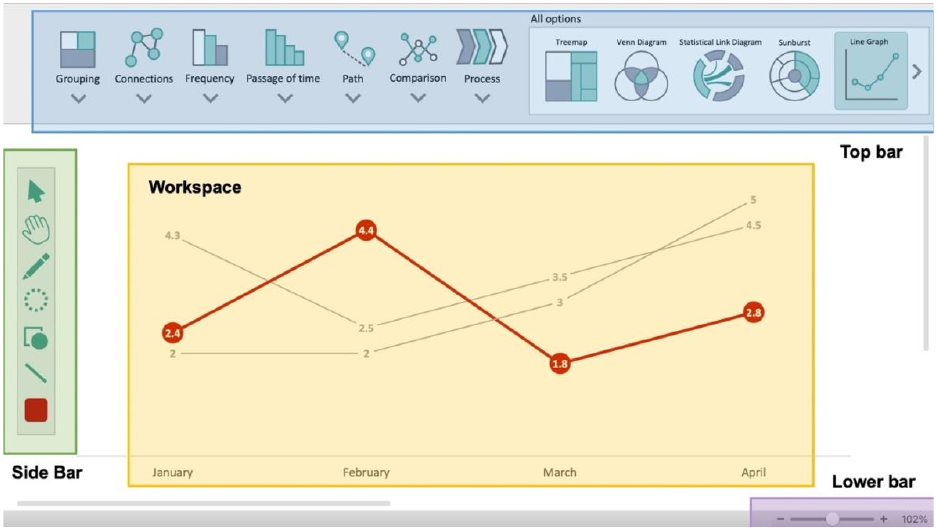}
    \caption{Our prototype interface. We used this prototype as a design probe, and discussed how it would support qualitative research with our participants.}
    \label{fig:prototype}
\end{figure}

\replaced{Following these guidelines, we developed a prototype (\autoref{fig:prototype})}{A prototype was developed}, which \replaced{comprised}{included} 16 pages of \added{paper-based} examples \deleted{of how the software would function and look like}. \deleted{Our final prototype  mirrored PowerPoint, based on a suggestion by Participant 1, who wanted the interface to be familiar and easy to use.} \added{The prototype was designed to be simple and emulate familiar software like PowerPoint and Microsoft Paint. It supports researchers who know what they want to get out of their data (i.e., whether they want to check for frequency or see a path) and then choose the data visualization they think best fits their needs out of the ones in that category.}

The prototype consists of a simple screen, divided into 3 sections. The \deleted{main white part at the} centre \added{area} is where the data visualizations are created and edited. The top bar is where the researcher can choose what kind of data visualization will be generated based on their data. And lastly, the left side bar \replaced{provides}{are the} editing tools. \deleted{The prototype was designed to be simple and emulate other more wildly spread software like PowerPoint and Microsoft Paint. The idea is that the researcher has to know what they want to get out of their data (i.e., whether they want to check for frequency or see a path) and then choose the data visualization they think best fits their needs out of the ones in that category. After a data visualization is selected, the software works with the data the researcher imported and creates a visualization.}  \replaced{We included a variety of data visualizations}{The data visualizations were added} based on \replaced{those identified}{the ones suggested} by \deleted{the} participants \added{in our interviews.} \deleted{and the ones they like to use the most when conducting thematic analysis. The categories were made to foster agency through choice, as the users need to choose what they want the machine to do for them. Additionally, they function as guidance for the user so they do not feel frustrated with the excess of freedom and lack of instruction. Both of these elements were in accordance to the participant's suggestions during the first interview.}

\deleted{The top part of the interface separated the data visualizations into the most common categories brought up by the participants, and an area for all data visualizations without separation by category. This allowed the participants organize their thoughts and understand the main roles of each visualization based on their own needs. This was done following comments from the participants that highlighted their need for data visualizations but lack of knowledge on what is available and how best utilize what exists. Because the participants showed proficiency in how they wanted to use visualizations, the categories were created to foster agency through choice and allow them to pick what they wanted out of an array of possibilities.}

\deleted{The data visualizations provided by the software are based on the ones the participants mentioned using the most when conducting Thematic Analysis, such as tables, venn diagrams, linking and grouping diagrams, and mind maps. }

\deleted{As per the requests from participant 1 the software was idealized to be malleable and for the user to be able to quickly change between different visualizations in order to see which one fits their needs best.} \added{The prototype draws on design language from Microsoft PowerPoint and NVIVO, which were both cited as tools used by the participants to create their visualizations. This was in line with the concepts of fostering agency through mastery of space, therefore we elected to make the changes and the navigation as seamless and simple as possible.}

\replaced{Lastly, the}{The} side panel to the left provides users with editing tools for tailoring their data visualizations to their specific needs. Editing tools were requested by \replaced{P}{p}articipants 1 and 3 who mentioned wanting some degree of creative freedom to make the data visualizations better fit their desired narratives. Additionally, the tool sought to foster agency through collective creation by enabling the user to work alongside the software.

\deleted{Lastly, movement of the software is intended to replicate Microsoft PowerPoint, and NVIVO, which were both cited as tools used by the participants to create their visualizations. This was done in an effort to make movement intuitive and not cause frustration.}



\section{Feedback on Prototype}

\added{We invited the same five participants from our initial interviews to return and provide feedback on our prototype. Participants were introduced to the prototype’s functionality for conducting thematic analysis, shown images, and provided with a ZOOM Whiteboard session where they could annotate freely to communicate their feedback and suggest modifications. We then performed a thematic analysis of their responses, consistent with the analysis we performed in the original round of interviews.}

\added{Overall, feedback was positive, with participants finding the semi-automated data visualization process promising for enhancing research workflows. They commented on two aspects that affected their interest in the prototype, and those were their autonomy, and visualization literacy.}


\deleted{The testing of the prototype was done with the same 5 participants from the previous interview to continue the user-centered design process and ensure that the conclusions the research team reached from the interviews matched with the needs and wants of the participants. The participants were shown images of the prototype and had the researcher walk them through how they could use it to conduct thematic analysis. Following that, they were presented with all pages of the prototype on a ZOOM Whiteboard and allowed to sketch over as they saw fit to better express their feelings about it or make any changes they desired.  Additionally, We asked questions about how they envisioned using the tool to conduct their own research, as well as asked to describe any additional wants, needs, or issues they might have had.}




\deleted{The feedback was generally positive, with participants voicing their belief that the software could help them with their research. Overall, the semi-automated process of generating data visualizations to help translate information between the researcher and the AI was considered to be a positive and incited curiosity in the participants for further work alongside AI.}

\deleted{Some visualizations scored lower in participant understanding, amongst them the statistical time map and the statistical linking diagram confused the participants the most. Statistical linking diagrams have the purpose of linking data points as well as showing their prevalence within a sample, therefore acting as a mixed-methods way of presenting data. However, participant 5 and 2 initially dismissed statistical linking diagrams as solely quantitative , only processing the qualitative aspect of it after some explaining by the interviewer.}

\deleted{Participant 5 quoted that she “never thought that [statistical linking diagrams] could be done like that. It's really nice,” upon understanding the applications of it for qualitative analysis. On the other hand, participant 4 had difficulties understanding the statistical time map, which is used to represent statistical changes of data across time and space. She explained she remembered what it was from the last interview, but that she would still consider it a complicated data visualization to be available with no explanation. She quoted “I think you explained the statistical time map to me. But I feel like if I don't know what it is, and then I press it, and let's say, let's say… vomits out a statistical time map I might not know what it is.”}

\deleted{On the other hand, participant 1 expressed liking the addition of those data visualizations, as they are the ones she did not feel confident she could do by herself, but would still be interested in including them in her research practices. She explained ``Because I've even had that need myself. And I'm not experienced. […] Like an easily accessible need, because even my committee members like ‘yeah, when I was a master student it was really hard to even figure out how to make a statistical linking diagram’.'' Both her and participant 5 shared that that AI could be very useful in these instances when frequency needs to be monitored, since machines can better keep track of numbers. They agreed that frequency checking is a task they are happy to delegate to AI in order to free their time for other parts of research.}

\deleted{Additionally, the participants asked for 6 data visualizations to be added for the prototype: hierarchy, Gantt charts, word clouds, word count, tabular grouping process, and nesting diagram. Additional feedback included: the addition of organization tabs, a text tool, a ruler on top of the drawing space, a grid, a color wheel, HEX code, eye-drop, heading for the categories of data visualization, and categories separating data visualizations into quantitative, qualitative, and mixed methods.}


\subsection{Autonomy, Delegation, and Transparency}

\replaced{Conversely}{Contrarily}, when it came to delegating tasks, participant 3 brought up the issue of transparency. She explained that working with machines can sometimes feel like interacting with a “black box” \replaced{that}{as it} gives the user results without disclosing the process. According to her, if a person is not well-versed in programming, and the machine delivers an unsatisfactory result, there is little guidance on how to alter the process to arrive at a satisfactory outcome. To \replaced{address}{solve} this issue, she suggested adding a feature that would \replaced{offer}{allow for} an explanation of how the AI has reached its results. This feature could \replaced{enhance}{combat that lack of} transparency by allowing the user to understand how a result was reached and \replaced{providing guidance for adjustments to the}{how to make changes to alter the} AI output. \deleted{She noted, “So, presenting them with some kind of clarity on like… why things were done might help to clear up some of that mental process and be able to like point to like one part of the process and be like, OK, like this is where we went wrong.”}

Similarly, participant 5 appreciated the transparency elements of the prototype, saying she is interested in \replaced{verifying}{checking} the results given by the AI and screening them. She compared it to “checking back math,” describing her approach to overseeing how the AI reached its conclusions and deciding whether to keep those results. Participant 2 echoed this sentiment, stating that the transparency feature in the prototype would be very useful if it \replaced{provided}{could explain to the user in} enough detail to allow for action. She noted, “I think that would make a lot of sense as long as it's detailed enough.”

Regarding positive examples of delegation, participants 1, 3, 4, and 5 \replaced{expressed appreciation for}{talked positively about} the choices that the tool allowed them. Participant 1 \replaced{appreciated}{expressed her appreciation of} the categories presented by the tool, which provided guidance without overstepping the boundaries of autonomy. She explained that having various options available, while still requiring the researcher to specify their needs, was an effective way to foster autonomy because it encourages the researcher to understand their data and \replaced{to clarify}{what they want to get out of} an analysis based on their objectives. She emphasized that through this sense of autonomy, the human aspect of qualitative research would \replaced{remain intact}{not get lost}.

\subsection{Visualization Literacy}

Additionally, participants 1, 3, 4, and 5 \replaced{appreciated having a}{talked about their appreciation for a} tool that could help them organize their thoughts in image form easily. Participant 3 shared that \replaced{she often knows}{sometimes she knows} what kind of visualization she wants but has a limited vocabulary to express those needs and \replaced{lacks the technical knowledge}{a lack of knowledge in how} to create the visual she has in mind. \replaced{A tool that presents her with visual options therefore facilitates her process of translating her mental image into a concrete visualization}{Therefore, having a tool that presents her with visual options facilitates the process of understanding how to bring to paper the image she has in her mind}. Similarly, Participant 4 expressed how useful it was to have a tool that \replaced{alleviates}{could remove} the burden of creating a visualization from scratch without guidance. She explained that having predefined categories \replaced{helps her make}{makes it easier for her to make} decisions. Participant 5 also \replaced{valued}{shared how she appreciates} a tool that makes her data analysis more efficient and allows her to explore connections more easily. She explained, “And I feel like a tool like this probably would have made the connections seem easier, like on my face. Which would have probably made my life easier.”

Participant 1 echoed similar sentiments, sharing that, for her quantitative research, she prefers software with pre-prepared data visualization options over software that requires her to know the exact type of visualization she wants and to do the coding to achieve it. \deleted{She noted, “I know it’s reliable and good… but it takes a long time and I have to learn how to do everything. […] But! I can’t edit it! I can only export it! I can probably edit it in the code, but that’s too hard. And then in contrast! I have this software, which… is so user-friendly.”}

Lastly, two participants used the low-fidelity prototype during the interview to experiment with ways of representing their own data. Participant 5, together with the interviewer, \replaced{browsed}{searched through} the categories and concluded that one of the tool’s options better suited her needs than her current visualization method. What had been a mind map now became a group linking diagram, which she felt more effectively demonstrated connections between elements belonging to two distinct groups.

Initially, Participant 5 expressed frustration with not even being able to sketch her idea, quoting, “Honestly, I don't know how to draw that out.” However, as she and the interviewer brainstormed potential visualization forms, her confidence grew until she arrived at a representation she was pleased with, stating, “Yeah, it makes more sense to me. Yeah, I like it. I feel like these are the things that makes you, I don't know, gives you ideas because you can actually build connections between two variables in your research more effectively.”

Overall, participants demonstrated \replaced{an understanding of}{prowess in} how each data visualization could be used and an awareness of how to choose the best visualization based on specific needs. Their ability to abstract data and envision how each representation could fulfill a different role \replaced{signalled that the interface should offer guidance but ultimately allow users to make the final choice}{signalled to us that that the design of the interface should be left with guidance, but still leave the ultimate choice up for the users}.

\section{Discussion}

Our analysis was motivated by a need to understand how qualitative researchers experience the four types of personal agency when using qualitative analysis software: \emph{action}, \emph{choice}, \emph{narrative}, and \emph{space}. We now discuss our findings related to each facet of personal agency, and how they help us understand how to design qualitative data visualization tools and interactions with ML and AI.

\subsection{Action}
The biggest barrier to fostering agency through action was delegation. The more tasks are delegated to the machine, the less choices the user has to make, and therefore there are less opportunities for agency through action to be fostered. 

Overall, participants were open to \emph{some} delegation, but were firm in maintaining their sense of autonomy and control. The tasks they were willing to delegate were the ones they considered to be time consuming and robotic, especially counting or checking for simple themes. However, participants were adamant about maintaining their autonomy on two fronts: double-checking the work performed by a machine and making choices for their own research.  

An issue raised by the participants when it came to AI assistance was their lack of trust in it. There were two sides to this mistrust, either participants did not believe AI would be competent enough to get useful results, or they feared AI would be too good and replace them. As a response to this lack of trust, an element that was appreciated in the prototype were the explanations. Participants reacted positively to having a tool that would allow them to work alongside AI by providing explanations about how the results were reached, commenting on how they felt the transparency made them feel more in charge.  

Explaining the results can contribute to establishing trust between user and AI \cite{Lubars2022}. In order for that trust to be cultivated, the people need to know the intend behind the automation and how it aligns with their own personal wants and needs, which can only work if the machine provides the user with enough information for them to understand it~\cite{Lubars2022}.

The explanations suggested for the prototype allowed the participants to either check back on what the AI had done to ensure the results were up to par with the needs of the researcher (e.g., participant 5 comparing it to checking back math), or gave them confidence to believe they were still in control of the research (e.g., participant 4 expressing her desires to remain the one to make choices and not have the AI take that autonomy from her). Participant 2 talked about that same feeling of wanting transparency, explaining that it would be ideal for her if the AI could explain what it had done with enough detail to allow her to make any changes if she wanted. Therefore, if the explanation is presented in a way that can allow for the user to understand the AI, then they can feel confident in acting upon those results, fostering agency. 

During the sample analysis, issues with fostering agency through action were found for the software that did not provide transparency, which meant the users had to spend significant time making sense of the results given to them by the AI. Participant 3 echoed this feeling when she shared how frustrating it could get to work with AI as a “black box” and have no idea where the results came from or how to change the query to get different ones. Because of this uncertainty, acting can become difficult and hinder agency.  

The use of the categories for data visualizations was similarly found to be positive in terms of fostering agency through action. Four out of five participants described using data visualizations as cognitive assistants when conducting research, which helped them organize, explore, and ultimately take control of their own research, which is in line with previous research \cite{Braga2020, Kennedy2020, Eichner2014, Cabitza2016}. The categories allowed them to visualize how to use their data and then experiment with it. This opportunity for action was praised by the participants who reported liking some level of guidance, but still being allowed to make the final choice themselves.

Lastly, through the editing tools provided in the interface, agency was fostered thorough the researchers’ interactions with the datavis \cite{Rawlins2014, Tanenbaum2009}. This feeling was being fostered through allowing the researchers’ opportunities to act by changing the datavis and seeing the changes in real time \cite{Murray1997, Rawlins2014, Eichner2014}. Participants reported noticing this by showing appreciation for the editing tools available and expressing how they liked being able to change the visualizations that were given by the machine.

\subsection{Choice}

Trust was the element that moderated agency through choice the most. Like the research done before \cite{Jiang2021, Feuston2021, Lubars2022} our participants expressed being open to working with machines as an assistant, but against full automation, citing they did not want their choices taken from them. The fact that our tool suggested possible results but ultimately left the choice up to the researchers was an element that participants commented improved their sense of perceived autonomy. 

Additionally, both the results from the interview and the analysis of the sample showed that another element that influences trust is receiving explanations for the actions of the AI. During the sample analysis, not providing transparency was considered to be an element that hindered the perception of agency, since it merely provided users with a result they could not interpret without further understanding the context the AI applied to it. Similar to action, when users do not understand how the AI reached the results, they do not feel confident in making choices, hindering their perception of agency. 


Elements that were considered to be positive were the ones that allowed participants to choose freely (i.e., editing tools and the categories of datavis). Chiefly amongst the elements that participants mentioned fostering agency were the categories and the ability to choose which datavis best suited their goals. There was an emphasis on wanting the effects of their choices to show up quickly. Participant 1 and 4 voiced that they would like the tool to adapt quickly when they made changes to the data so they could explore and immediately see if the changes they made were desirable. All these elements contribute to the idea that participants perceive agency when they make choices and see their effect on the world around them \cite{Murray1997}. Because of the multiple categories participants felt comfortable choosing the one that best suit them and quoted that possibility of choosing as a net positive. 

During the sample analysis, an element that was considered a barrier for choice was not allowing a user to make changes or input new elements. Two of the software analyzed failed in that aspect, restricting the users from creating new codes or editing the data visualizations. These features are the opposite of what participant 1 and 2 talked about wanting in a tool for assisting with research. Both shared enjoying working with AI the most when it provided suggestions and allowed them the option to accept or reject, providing opportunities for fostering agency through choice \cite{Eichner2014}. Furthermore, participants showed different preferences in the ways they used visualizations to process information. Because of these differences, they were receptive to a tool that catered to their many different needs and provided them with multiple choices to pick from. 

However, despite participants appreciating the freedom of choices, they also expressed a need for a tool that provides guidance. Participants quoted liking not having to entirely come up with the visualization themselves, instead delegating that task to the machine and only editing the results given. Additionally, the categories were noted as a positive as they provided them enough guidance to make a choice without completely removing autonomy. This combination of guidance and editing freedom contributed to fostering agency, which is in line with previous literature that point out that in order to perceive agency through making choices restrictions are necessary \cite{Eichner2014, Tanenbaum2009}. Participant 1 illustrated this point when she mentioned preferring software that gives her guidance, instead of the one with complete freedom, but no assistance. She mentioned how she would rather use the one that was more restricting, than struggling through figuring out what she wants from scratch with no help from the system. 


\subsection{Narrative}

Mastery of narrative occurs when people detect the genre of the media they are interacting with and can accurately predict what will happen based on previous experiences \cite{Eichner2014}. During our interviews, participants frequently mentioned using data visualization for narrative purposes, both for finding the story in their own data and to weave a story for others.  Participant 3 highlighted the importance of datavis when communicating information and the need to focus on narrative to garner tangible change. Using datavis as a way to engage people and promote change has been discussed before, as it is a type of media that can increase intrinsic motivation and empathy \cite{Braga2020,Gray2020,Cooley2020}. 

However, the prototype also fostered agency through narrative by mutual creation \cite{Rawlins2014}. Participants expressed liking the editing tools and flexibility of a software that allows them to craft the datavis to best fit their desired narrative \cite{Rawlins2014,Ryan2006}. Another way to look at the positive reaction to editing tools is through a commitment to meaning \cite{Tanenbaum2009} and the crafting of said meaning by participating in creation \cite{Rawlins2014}.  

Especially when it comes to crafting a narrative, participants talked about liking the use of data visualizations, sharing how they prefer to let the data lead the narrative and create a story that makes sense using visual elements. Participant 5 shared how she prefers to visualize data before crafting a story, and that is the way she found makes the most sense for her way of thinking. The types of data visualization most appreciated for narrative purposes were the ones that showed relationships and connections. Similarly, the participants commented that through mapping out connections they had an easier time following the thread of a story, which is in line with \citet{Eichner2014} and the idea that by recognizing similarities in media, a user can follow a narrative and experience agency \cite{Eichner2014}. 


\subsection{Space}


Overall, participants enjoyed using data visualizations to conduct data analysis, reporting that the act of visualizing things made it easier for cognitive processes. However, a caveat of those visualizations was the need for freedom of movement within that visualization. 

Participants reported feeling frustration with tools that had complicated navigation systems and did not allow them to move freely. The less intuitive it is to move through a digital space, the less agency through space is fostered \cite{Murray1997, Bizzocchi2003}. Additionally, participant 4 also commented preferring to use whiteboards, paper, or her iPad when making data visualizations for organizing her thoughts, as they gave her the freedom of movement for being messy. Similarly, participant 2 and 3 also described the act of physically moving elements around as a positive and appreciated it during research. All these actions of moving through digital space work for fostering agency through the mastery of space \cite{Murray1997}. 

During the sample analysis, similar results were found, in which the software that allowed for movement were considered better for fostering agency through movement. For example, NVIVO’s ability to click on elements and move them around freely mimics the experience participant 4 described when she talked about being able to pick elements up and move them around during data exploration. Another example of mastery of space interfering with perceptions of agency comes from participant 3, who discusses her frustration in trying to navigate word files as they had no features to facilitate moving in a way that is needed for data analysis.

Lastly, guidance was regarded as an important part of navigation during the sample analysis. For example, NVIVO offers options for further analysis, but they are hidden under right-clicks that the user either has to discover or already know. This lack of guidance can be considered a barrier since it makes navigation confusing and frustrating \cite{Eichner2014, Murray1997}. 

Participant 3 reported a similar frustration with her use of the CTA \cite{Gauthier2022}, quoting that the lack of guidance made navigating the software difficult. Both the participants and the sample analysis showed that it is important to have proper guidance when trying to foster agency through the mastery of space, since the more frustration navigation causes the less agency is perceived \cite{Bizzocchi2003, Eichner2014}.

Other elements were reported by the participants as helping with navigation, such as the use of colour-coding and clustering. According to participant 4, the use of colours allows her to organize data in her brain. In terms of clustering, the participants enjoyed the separated sections of the prototype, sharing that it made it easy to navigate and get to where they wanted from the main screen.


\section{Implications for HCI}

Our work provides three sets of implications for HCI research moving forward, which we now reflect on.

\subsection{Agency as Design Lens}

Human-computer interaction research has a long-standing need to better understand how to engage and motivate \cite{Bennett2023}. Self Determination Theory \cite{deci2012self}, for instance, is already \deleted{a} well-established \deleted{theory} in HCI research (e.g. \cite{Tyack2020,CHISDTWorkshop}) with core concepts that closely parallel personal agency. \citet{Tyack2020} calls for deeper engagement with the theory, gaps in our understanding of how it should be applied to interface design, and point to the need for future research in this area. We assert that personal agency provides a practical lens through which to think about motivation and engagement in user-centred design.

Subtle differences in each theory's core concepts may provide important context to designers.  SDT's concept of \emph{autonomy} refers to the capacity to make independent decisions and act based on one's own volition, which emphasizes independence and self-governance. \emph{Agency}, on the other hand, is the ability to act in a given environment and influence it, focusing on the power to enact change. Both are related but not synonymous.

A notable challenge in adopting SDT has been its abstractness. That is, while the concepts of autonomy, relatedness, and competence are powerful they do not map well to design thinking. We argue that personal agency more closely aligns with design work since the concepts of narrative, action, choice, and space can be more easily translated into tangible changes. By following the lens of personal agency, it allowed us to map out how each design elements fostered agency and how the participants preferred it to be included in the prototype. Therefore, using personal agency as a design lens was useful for creating a prototype that fostered agency for researcher regarding their own research. This is especially useful when they were dealing with AI, which was previously found to take away from agency.

In the case of our prototype, some of the key design elements that were added to foster agency were: the categories for data visualization and the editing tools. The categories were added to foster agency through choice and action. By giving the users guidance on how they could use datavis for their research, but ultimately leaving it up to them how to employ them, we sought to foster an environment where individual choices and actions were preserved. As for the editing tools, they were added to foster agency through action and narrative, as they allowed the users to act upon a datavis and alter it to fit a specific narrative, inciting agency through mutual collaboration.

Finally, we point to the need for future work that more deeply engages with the concepts of personal agency and their application in HCI research. In this work we focused on personal agency, but creative and collective agency are related concepts that may also provide a useful lens for design. Similarly, collective agency parallels the SDT concepts of connectedness and relatedness, and like our exploration here, may provide a means of more deeply engaging in those aspects of design. 




\subsection{Motivating More Interactive ML}

Our findings parallel those of \citet{Jiang2021} and \citet{Feuston2021} where agency was considered important for qualitative researchers, and ownership over one's research was paramount. Much of what our participants expressed during their interviews was simply a need for \emph{agency} in the process: the ability to act and have influence on what has happened. They spoke about their desire to edit AI outputs, especially with the assistance of a visual interface. None of our participants were well versed in coding and they expressed that they felt like they could not alter or control the outputs of the AI unless they had a visual interface to assist them. 

This revealed to us a need participants had for interactive ML and for them to be able to control the AI even if they were not well versed in the coding behind it. The addition of explanations to make the AI process more transparent for non-AI experts was shown to be a key feature for instilling more trust in machine process and thus more openness towards task delegation. This follows the findings by \citet{Jiang2021} where they cite that AI should be employed as a collaborator for qualitative researchers, but never take their place.

Despite this collaboration seeming complicated, our results suggest that while, initially, qualitative researchers may be hesitant to adopt ML techniques in their work, the availability of a visual interface and transparency in the AI process is necessary for them to feel comfortable enough to consider delegating some tasks to the machine. \citet{Lubars2022} discusses this phenomenon when they talk about trust and delegation and how people can only trust machines to do their work when they understand the process happening behind the code. Therefore, the inclusion of transparency features in any AI tool focused on qualitative research is imperative.




\subsection{Design Process as Intervention}

Finally, participant perceptions of AI changed through the course of our study, highlighting the potential of collaborative design to facilitate buy-in from skeptical or hesitant qualitative researchers. During the first interview, 3 participants expressed that they were reluctant to work with AI, but after the co-design session, they were much more open to the idea of AI delegation. We assert that the co-design process helped participants to understand how collaboration with AI could be possible and reassured them that the software would not replace them, but might help with their research. 

This process of co-design functioned as an intervention to make participants more open and interested in using AI with their research methods. This is expected to happen when using user-centred design methods, as the participants become more used to the idea as they work on the solution alongside the designers. These results are in line with previous research \cite{Jiang2021,Feuston2021} which shows that qualitative researchers are open to AI assistance as long as they can understand and control how the work is being delegated to the AI. Through participating in the design of the prototype, the participants also understood the role of AI in their research and built trust in the machine \cite{Lubars2022}, and thus became more open to collaboration.

For the future, we would like to highlight the importance of designing with qualitative researchers when making tools to assist for their research. Not only did this process assist us in creating a tool they felt comfortable using, but it also served to build trust between them and the machine they would work with in the future. The use of DSR, an iterative method, should be further explored for HCI, not only as a method for the creation of artifacts and knowledge, but also as an intervention for users to become more comfortable with possible interventions.

\section{Conclusion}

As qualitative researchers seek to engage with ever-larger data sets, there is a need to understand how tools like ML and AI can support \replaced{their}{those} research activities. Our work shows several key considerations in developing those tools: the researchers' knowledge, skills, and awareness of computer visualization, their acceptance of ML supports, and an awareness of the benefits that ML can provide. We have also shown how the concepts of agency can be used to explore this design space, and provide insight into how ML interfaces can be designed for qualitative researchers moving forward.

\bibliographystyle{ACM-Reference-Format}
\bibliography{library_Luka_thesis}










\end{document}